\documentclass[11pt,epsf]{article}
\usepackage{amsmath}
\usepackage{amsfonts}
\usepackage{amssymb}
\usepackage{graphicx}
\usepackage{color}
\newtheorem{theorem}{Theorem}

\newcommand {\eqd} {\stackrel{\Delta} {=}}

\newcommand {\reals} {{\rm I\!R}}

\newcommand {\bu} {\mbox{\boldmath $u$}}
\newcommand {\bv} {\mbox{\boldmath $v$}}

\newcommand {\bx} {\mbox{\boldmath $x$}}
\newcommand {\by} {\mbox{\boldmath $y$}}
\newcommand {\bz} {\mbox{\boldmath $z$}}

\newcommand {\bE} {\mbox{\boldmath $E$}}

\newcommand {\bU} {\mbox{\boldmath $U$}}
\newcommand {\bV} {\mbox{\boldmath $V$}}

\newcommand {\bX} {\mbox{\boldmath $X$}}

\newcommand {\bZ} {\mbox{\boldmath $Z$}}

\newcommand{\calP}{{\cal P}}

\newcommand{\calS}{{\cal S}}

\newcommand{\calU}{{\cal U}}
\newcommand{\calV}{{\cal V}}

\newcommand{\calX}{{\cal X}}

\begin{document}
\thispagestyle{empty}
\title{On Optimum Strategies for Minimizing the Exponential Moments of a
Given Cost Function
}
\author{Neri Merhav
}
\date{}
\maketitle

\begin{center}
Department of Electrical Engineering \\
Technion -- Israel Institute of Technology \\
Technion City, Haifa 32000, ISRAEL\\
{\tt merhav@ee.technion.ac.il}\\
\end{center}
\vspace{1.5\baselineskip}
\setlength{\baselineskip}{1.5\baselineskip}

\begin{abstract}
We consider a general problem of finding 
a strategy that minimizes the exponential moment of a given cost
function, with an emphasis on its relation to the more common criterion of
minimization the expectation of the first moment of the same cost function.
In particular, our main result is a theorem that
gives simple sufficient conditions for a strategy to be optimum
in the exponential moment sense. This theorem may be useful in various
situations, and application examples are given. We also examine the asymptotic
regime and investigate universal asymptotically optimum strategies in light of
the aforementioned sufficient conditions, as well as phenomena of
irregularities, or phase transitions, in the behavior of the asymptotic
performance, which can be viewed and understood from a statistical--mechanical
perspective. Finally, we propose a new route for deriving lower bounds on
exponential moments of certain cost functions (like the square error in
estimation problems) on the basis of well known lower bounds on their
expectations.\\

\vspace{0.2cm}

\end{abstract}

\section{Introduction}

Many problems in information theory, communications, statistical signal
processing, and related disciplines can be formalized as being
about the quest for a strategy $s$ that 
minimizes (or maximizes) the expectation 
of a certain cost function, $\ell(X,s)$, where
$X$ is a random variable (or a random vector). 
Just a few examples of this generic paradigm are the following: (i)
Lossless and lossy data compression, where $X$ symbolizes the data to be
compressed, $s$ is the data compression scheme, and $\ell(X,s)$ is the length
of the compressed binary representation, or the distortion (in the lossy
case) or a linear combination of both (see, e.g., 
\cite[Chapters 5 and 10]{CT06}).
(ii) Gambling and portfolio theory \cite[Chapters 6 and 16]{CT06}, where cost
function is logarithm of the wealth relative.
(iii) Lossy joint source--channel coding, 
where $X$ collectively symbolizes the randomness of  
source and the channel, $s$ is the encoding--decoding scheme and $\ell(X,s)$
is the distortion in the reconstruction (see, e.g.,
\cite{Wyner80},\cite{Wyner81}). (iv)
Bayesian estimation of a random variable based on measurements, where $X$
designates jointly the desired random variable and the measurements, $s$ is
the estimation function and $\ell(X,s)$ is the error function, for the
example, the squared error. Non--Bayesian estimation problems can be
considered similarly (see, e.g., \cite{VanTrees68}). (v)
Prediction, sequential decision problems (see, for example, 
\cite{MF98}) and stochastic control problems \cite{Bertsekas07}, such as the linear
quadratic Gaussian (LQG) problem, as well as general Markov decision
processes, are also
formalized in terms of selecting strategies in order to minimize
the expectation of a certain loss function.

While the criterion of minimizing the expected value of $\ell(X,s)$ has been
predominantly the most common one, the exponential moments of $\ell(X,s)$,
namely, $\bE\exp\{\alpha\ell(X,s)\}$ ($\alpha > 0$), have received much less
attention than they probably deserve in this context.
There are a few motivations for
examining strategies that minimize exponential moments. First,
$\bE\exp\{\alpha\ell(X,s)\}$, as a function of $\alpha$, is obviously the
moment--generating function of $\ell(X,s)$, and as such, it provides the full
information about the entire distribution of this random variable, not just its first
order moment. Thus, in particular, if we are fortunate enough to find a strategy that
uniformly minimizes $\bE
\exp\{\alpha\ell(X,s)\}$ for all $\alpha \ge 0$ (and there are examples
that this may be the case), then this is much stronger than just minimizing the
first moment. Secondly, exponential moments are intimately related to
large--deviations rate functions, and so, the minimization of exponential
moments may give us an edge on minimizing probabilities of (undesired) large deviations
events of the form $\mbox{Pr}\{\ell(X,s) \ge L_0\}$ (for some threshold $L_0$), or more precisely,
on maximizing the exponential rate of decay of these probabilities. There are
several works along this line, especially in contexts related to 
buffer overflow in data
compression \cite{FM86},\cite{Humblet81},
\cite{Jelinek68},\cite{Merhav91},\cite{MN92},\cite{UH99},\cite{Wyner74-1},
and exponential moments related to guessing \cite{Arikan96},\cite{AM98a},
\cite{AM98b},\cite{Massey94},\cite{MA99},\cite{MRA99}.

It is natural to ask, in view of the foregoing discussion, how can we harness
the existing body of knowledge concerning optimization of strategies for
minimizing the first moment of $\ell(X,s)$, which is quite mature in many
applications, in our quest for optimum strategies that minimize exponential
moments. Our main basic result, in this paper, is a simple theorem that
relates the two criteria. In particular, we 
furnish sufficient conditions that 
the optimum strategy in the exponential moment sense can be
found in terms of the optimum strategy in the first moment sense, for a
possibly different probability distribution, which our theorem characterizes.

In some applications, these
sufficient conditions for optimality
in the exponential moment sense, yield an equation in $s$, whose solution is
the desired optimum strategy. It is clear then that in these applications,
the optimality conditions provide a concrete tool for deriving the optimum
solution. In other applications, however, this may not be quite the case
directly, yet the set of optimality conditions may still
serve as a useful tool: More often than not, in a given instance of the problem
under discussion, one may have a natural intuitive
guess concerning the optimum strategy, and then the optimality conditions
can be used to prove 
that this is the case. 
One example for this, that will be demonstrated in detail later on, is the following: 
Given $n$ independent and identically distributed 
(i.i.d.) Gaussian observations, $X_1,\ldots,X_n$, with mean
$\theta$, the sample mean, $s(X_1,\ldots,X_n)=\frac{1}{n}\sum_{i=1}^nX_i$,
is the optimum unbiased estimator of $\theta$, not merely in the
mean squared error sense (as is well known), but also in the sense of minimizing all
exponential moments of the squared error, i.e.,
$\bE\exp\{\alpha[s(X_1,\ldots,X_n)-\theta]^2\}$ for all $\alpha \ge 0$ for which
this expectation is finite.

We next devote some attention to the asymptotic regime. Consider the case where
$X$ is a random vector of dimension $n$, $X=(X_1,\ldots,X_n)$, governed by
a product--form probability distribution, and $\ell(X,s)$ grows linearly for a
given empirical distribution of $X$, for example, when
$\ell(X,s)$ is additive, i.e., $\ell(X,s)=\sum_{i=1}^n l(X_i,s)$. In this case, 
the exponential moments of $\ell(X,s)$ typically behave (at
least asymptotically) like exponential functions of $n$. 
If we can then select a strategy $s$ that somehow ``adapts''\footnote{
The precise meaning of this will be
clarified in the sequel.}
to the empirical
distribution of $(X_1,\ldots,X_n)$, then such strategies may be universally
optimum (or asymptotically optimum in the sense of achieving the minimum
exponential rate of the exponential moment) in that they depend on neither the underlying probability
distribution, nor on the parameter $\alpha$. This is demonstrated in several
examples, one of which is an extension of a well known result by Rissanen in
universal data compression \cite{Rissanen84}.

An interesting byproduct of the use of the exponential moment criterion in
the asymptotic regime is
the possible existence of phase transitions: In turns out that the
asymptotic exponential rate of $\bE\exp\{\alpha\ell(X_1,\ldots,X_n,s)\}$ as a
function of $n$, may not be a smooth function of $\alpha$ and/or the parameters
of the underlying probability distribution even when the model under
discussion seems rather simple and `innocent.'
This is best understood from a
statistical--mechanical perspective, because in some cases, the calculation of
the exponential moment is clearly analogous to that of the partition function
of a certain physical system of interacting particles, which is known to
exhibit phase transitions. It is demonstrated that at least in certain cases, these phase
transitions are not merely an artifact of a badly chosen strategy, but they
appear even when the optimum strategy is used, and hence these phase
transitions are inherent in the model.

We end this paper by touching upon yet another aspect of the exponential
moment criterion, which we do not investigate
very thoroughly here, but we believe it is interesting and therefore
certainly deserves a further study in the future:
Even in the ordinary setting, of seeking strategies that minimize
$\bE\{\ell(X,s)\}$, optimum strategies may not always be known, and then
lower bounds are of considerable importance as a reference performance figure.
This is {\it a--fortiori} the case when exponential moments are considered. One
way to obtain non--trivial
bounds on exponential moments is via lower bounds the expectation of
$\ell(X,s)$, using the techniques developed in this paper.
We demonstrate this idea in the context of a lower bound on
the expected
exponentiated squared error of an unbiased 
parameter estimator, on the basis of the
Cram\'er--Rao bound (CRB),
but it should be understood that, more generally, the same idea can be applied
on the basis of
other well--known bounds of the mean-square error (Bayesian and non--Bayesian)
in parameter estimation, and in
signal estimation,
as well as in other problem areas.

\section{Basic Optimality Conditions}

Let $X$ be a random variable taking on values in a certain alphabet $\calX$,
and drawn according to a given probability distribution $P$. Let the variable
$s$ designate a {\it
strategy} chosen from some space $\calS$ of allowed strategies. The term
``strategy'' in our context is fairly generic: it may be
a scalar variable, a vector, an infinite sequence, a function (of $X$), a
partition of $\calX$, a coding scheme for $X$, and so on. 
Associated with each $x\in\calX$ and $s\in\calS$, is
a loss $\ell(x,s)$. The function $\ell(x,s)$ is called the {\it loss function},
or the {\it cost function}.
The operator $\bE\{\cdot\}$ will
be understood as the expectation operator with respect to (w.r.t.) the
underlying distribution $P$, and whenever we refer to the expectation w.r.t.\
another probability distribution, say, $Q$, we use the notation
$\bE_Q\{\cdot\}$. Nonetheless, occasionally, when there is more than one probability
distribution playing a role at the same time and we wish to emphasize that the expectation is
taken w.r.t.\ $P$, then to avoid confusion, we may denote this expectation
by $\bE_P\{\cdot\}$.

For a given $\alpha > 0$, consider the problem of minimizing
$\bE\exp\{\alpha\ell(X,s)\}$ across $s\in\calS$. The following theorem
relates the optimum $s$ for this problem to the optimum $s$ for the problem
of minimizing $\bE_Q\{\ell(X,s)\}$ w.r.t.\ another probability distribution
$Q$.

\begin{theorem}
Assume that there exists a strategy $s\in\calS$ for which
\begin{equation}
Z(s)\eqd\bE_P\exp\{\alpha\ell(X,s)\} < \infty.
\end{equation}
A strategy $s\in\calS$ 
minimizes $\bE_P\exp\{\alpha\ell(X,s)\}$ if there exists
a probability distribution $Q$ on $\calX$ that satisfies the following two
conditions at the same time:
\begin{enumerate}
\item The strategy $s$ minimizes $\bE_Q\{\ell(X,s)\}$ over $\calS$.
\item The probability distribution $Q$ is given by
\begin{equation}
Q(x)=\frac{P(x)e^{\alpha\ell(x,s)}}{Z(s)}.
\end{equation}
\end{enumerate}
\end{theorem}
An equivalent formulation of Theorem 1 is the following: denoting
by $s_Q$ a strategy that minimizes $\bE_Q\{\ell(X,s)\}$ over $\calS$,
then the theorem asserts that $s_Q$ minimizes $\bE_P\exp\{\alpha\ell(X,s)\}$
over $\calS$ if 
\begin{equation}
\label{cond}
Q(x) \propto P(x)e^{\alpha\ell(x,s_Q)},
\end{equation}
where by $A(x)\propto B(x)$, we mean that $A(x)/B(x)$ is a constant,
independent of $x$.

\vspace{0.2cm}

\noindent
{\it Proof.}
Let $s\in\calS$ be arbitrary and let $(s^*,Q^*)$ satisfy conditions 1 and 2 of
Theorem 1. Consider the following chain of inequalities:
\begin{eqnarray}
\label{chain}
\bE_P\exp\{\alpha\ell(X,s)\} &=& 
\bE_{Q^*}\exp\left\{\alpha\ell(X,s)+\ln\frac{P(X)}{Q^*(X)}\right\}\nonumber\\
&\ge&\exp\left\{\alpha\bE_{Q^*}\ell(X,s)-D(Q^*\|P)\right\}\nonumber\\
&\ge&\exp\left\{\alpha\bE_{Q^*}\ell(X,s^*)-D(Q^*\|P)\right\}\nonumber\\
&=&\exp\left\{\alpha\bE_{Q^*}\ell(X,s^*)-
\bE_{Q^*}\ln\frac{e^{\alpha\ell(X,s^*)}}{Z(s^*)}\right\}\nonumber\\
&=& Z(s^*) =
\bE_P\exp\{\alpha\ell(X,s^*)\},
\end{eqnarray}
where the first equality results from a change of measure (multiplying
and dividing $e^{\alpha\ell(X,s)}$ by $Q^*(X)$), the second line is by
Jensen's inequality and the convexity of the exponential function (with
$D(Q\|P)\eqd\bE_Q\ln[Q(X)/P(X)]$ being the relative entropy between $Q$ and
$P$), the third line is by condition 1 of Theorem 1, and the remaining
equalities result from condition 2: On substituting 
$Q^*(x)=P(x)e^{\alpha\ell(x,s^*)}/Z(s^*)$ into $D(Q^*\|P)$, one readily
obtains $D(Q^*\|P)=\alpha\bE_{Q^*}\ell(X,s^*)-\ln Z(s^*)$.
This completes the proof of Theorem 1.
$\Box$ 

Observe that for a given $s$, Jensen's inequality in the second line
of (\ref{chain}), becomes an equality for $Q(x)=P(x)e^{\alpha\ell(x,s)}/Z(s)$,
since for this choice of $Q$, the random variable that appears in the
exponent, $\alpha\ell(X,s)+\ln\frac{P(X)}{Q(X)}$, becomes degenerate
(constant with probability one). Since the original expression is independent
of $Q$, such an equality in Jensen's inequality means that
$\alpha\bE_{Q}\ell(X,s)-D(Q\|P)$ is maximized by this choice of $Q$, a fact
which can also be seen from a direct maximization of this expression using
standard methods.
Thus, we have a simple identity for every $s$:
\begin{equation}
\label{identity}
\bE_P\exp\{\alpha\ell(X,s)\}=\exp\{\alpha\max_Q[\bE_Q\ell(X,s)-D(Q\|P)]\}.
\end{equation}
This identity will prove useful in several places throughout the sequel.

Suppose next that the set $\calS$ and the loss function $\ell(x,s)$
are such that:
\begin{equation}
\label{minimax}
\min_{s\in\calS}\max_Q[\alpha\bE_Q\ell(X,s)-D(Q\|P)]=
\max_Q\min_{s\in\calS}[\alpha\bE_Q\ell(X,s)-D(Q\|P)].
\end{equation}
This equality between the min--max and the max--min means that there is a
saddle point $(s^*,Q^*)$, where $s^*$ is a solution of the min--max
problem on the left--hand side and $Q^*$ is a solution to the max--min problem
on the right--hand side. It is easy to check 
that the maximizing $Q$ in the inner
maximization on the left--hand side is 
$Q^*(x)=P(x)e^{\alpha\ell(x,s^*)}/Z(s^*)$, which is
condition 2 of Theorem 1. By the same token,
the inner minimization over $s$ on the right--hand side obviously
minimizes $\bE_{Q^*}\ell(X,s)$, which is condition 1. It follows then
that if the min--max and the max--min are equal, then the saddle point
satisfies the conditions of Theorem 1, and hence the corresponding
$s^*$ is optimum. Note also that when eq.\ (\ref{minimax}) holds, the
conditions of Theorem 1 become also necessary conditions for optimality:
Suppose that $s^*$ is optimum. Then, by eq.\ (\ref{identity}), it must
solve the minimax problem on the left--hand side of eq.\ (\ref{minimax}).
But if eq.\ (\ref{minimax}) holds then there is a saddle point, 
and $s^*$ if the first coordinate
of this saddle point, $(s^*,Q^*)$. But then $s^*$ and $Q^*$ must be related
according to the conditions of Theorem 1, as explained above.

When does eq.\ (\ref{minimax}) hold? In general, the well--known sufficient
conditions for 
\begin{equation}
\min_{u\in\calU}\max_{v\in\calV}f(u,v)=\max_{v\in\calV}\min_{u\in\calU}f(u,v)
\end{equation}
are that $\calU$ and $\calV$ are convex sets (with $\calU$ being independent
of $v$ and $\calV$ being independent of $u$), and that $f$ is convex in $u$ and concave
in $v$. In our case, 
since the function $f(s,Q)=\alpha\bE_Q\ell(X,s)-D(Q\|P)$ is always concave in
$Q$, this sufficient condition would automatically hold whenever 
$\ell(x,s)$ is convex in $s$ (for every fixed $x$),
provided that $\calS$ is a space in which
convex combinations can be well defined, and that $\calS$ is a convex set.

\vspace{0.2cm}

\noindent
{\it Maximizing Negative Exponential Moments.}
A similar, but somewhat different, criterion pertaining to exponential
moments, which is reasonable to the same extent, 
is the dual problem of $\max_{s\in\calS}\bE\exp
\{-\alpha\ell(X,s)\}$ (again, with $\alpha > 0$). If $\ell(x,s)$
is non-negative for all $x$ and $s$, this has the 
advantage that the exponential moment
is finite for all $\alpha > 0$, as opposed to
$\bE\exp\{\alpha\ell(X,s)\}$ which, in many cases, is finite only for 
a limited range of $\alpha$. For the same considerations as before, here
we have:
\begin{eqnarray}
\max_s\bE\exp\{-\alpha\ell(X,s)\}&=&
\max_s\exp\{\max_Q[-\alpha\bE_Q\ell(X,s)-D(Q\|P)]\}\nonumber\\
&=&\exp\{-\min_s\min_Q[\alpha\bE_Q\ell(X,s)+D(Q\|P)]\},
\end{eqnarray}
and so the optimality conditions relating $s$ and $Q$ are similar to those
of Theorem 1 (with $\alpha$ replaced by $-\alpha$), except that now we have
a double minimization problem rather than a min--max problem. However, it
should be noted that here
the conditions of Theorem 1 are only
necessary conditions, as for the above equalities to hold, 
the pair $(s,Q)$ should {\it globally} minimize
the function $[\alpha\bE_Q\ell(X,s)+D(Q\|P)]$, unlike the earlier
case, where only a saddle point was sought.\footnote{In other 
words, it is not enough
now that $s$ and $Q$ are in `equilibrium' in the sense that $s$
is a minimizer for a given $Q$ and vice versa.} On the other hand,
another advantage of this criterion, is that even if one cannot solve
explicitly the equation for the optimum $s$, 
then the double minimization naturally suggests
an iterative algorithm: starting from an initial guess $s_0\in\calS$, one
computes $Q_0(x)\propto P(x)\exp\{-\alpha\ell(x,s_0)\}$ (which minimizes
$[\alpha\bE_Q\ell(X,s)+D(Q\|P)]$ over $\{Q\}$), then one finds
$s_1=\mbox{arg}\min_{s\in\calS}\bE_{Q_0}\{\ell(X,s)\}$, and so on. It is
obvious that $\bE\exp\{-\alpha\ell(X,s_i)\}$, $i=0,1,2,\ldots$, 
increases (and hence improves)
from iteration to iteration. This is different from the min--max situation
we encountered earlier, where successive improvements are not guaranteed.

\section{A Few Examples}

Theorem 1 tells us that if we are fortunate enough to find a strategy
$s\in\calS$ and a probability distribution $Q$, which are `matched'
to one another (in the sense defined by the above conditions), then we
have solved the problem of minimizing the exponential moment. Sometimes it is
fairly easy to find such a pair $(s,Q)$ by solving an equation. In other
cases, there might be a natural guess for the optimum $s$, which can be
proven optimum by checking the conditions. 
In this section, we will see examples of both types.
Some of these examples could have been also solved directly, without using
Theorem 1, but for others, this does not seem to be a trivial task.
In some of the examples, it turns out that the same optimum strategy that
minimizes expected loss, is also optimum in the sense of minimizing all
exponential moments, but this is, of course, not always the case.

\subsection{Example 1: Lossless Data Compression}

We begin with a very simple example.
Let $X$ be a random variable taking on values in a finite alphabet $\calX$,
let $s$ be a probability distribution on $\calX$, i.e., a vector
$\{s(x),~x\in\calX\}$ with $\sum_{x\in\calX}s(x)=1$ and $s(x)\ge 0$ for all
$x\in\calX$, and let $\ell(x,s)\eqd -\ln s(x)$. This example is clearly
motivated by lossless data compression, as $-\ln s(x)$ is the length function
(in nats) pertaining to a uniquely decodable code that is induced by a distribution $s$,
ignoring integer length constraints. In this problem, one readily observes
that the optimum $s$ for minimizing $\bE_Q\{-\ln s(X)\}$ is $s_Q=Q$. Thus, by
eq.\ (\ref{cond}), we seek a distribution $Q$ such that
\begin{equation}
Q(x) \propto P(x)\exp\{-\alpha\ln Q(x)\}=\frac{P(x)}{[Q(x)]^\alpha}
\end{equation}
which means $[Q(x)]^{1+\alpha}\propto P(x)$, or equivalently,
$Q(x)\propto [P(x)]^{1/(1+\alpha)}$. More precisely,
\begin{equation}
s_Q(x)=Q(x)=\frac{[P(x)]^{1/(1+\alpha)}}{\sum_{x'\in\calX}[P(x')]^{1/(1+\alpha)}},
\end{equation}
and the expectation of $-\ln s_Q(X)$
yields the R\'enyi entropy. Note that here
$\ell(x,s)$ is convex in $s$ 
and so, the minimax condition holds.
While this result is well known and it could have been obtained
even without using Theorem 1, our purpose in this example was to show
how Theorem 1 gives the desired solution even more easily than with the direct
method, by solving 
a very simple equation.

\subsection{Example 2: Bayesian Estimation}

Let $(X,Y)$ be random variables, where $Y$ is distributed according to a
given density $P(y)$ and the conditional density of $X$ given $Y$ is given by
\begin{equation}
P(x|y)=\frac{1}{\sqrt{2\pi}}\exp\left\{-\frac{1}{2}(x-\phi(y))^2\right\},
\end{equation}
where $\phi(y)$ is a given function. We are seeking the optimum linear
estimator of $X$ based on the observation $Y$ in the sense of minimizing
the exponential moment of squared error. In other words, we seek a real number
$s$ that minimizes
$\bE\exp\{\alpha(X-sY)^2\}$, where $\alpha\in(0,1/2)$.
Once again, the loss function is convex in $s$.
According to the second condition of Theorem 1, $Q$ should be of the form
\begin{eqnarray}
Q(x,y)&\propto&P(y)\cdot
\frac{1}{\sqrt{2\pi}}\exp\left\{-\frac{1}{2}(x-\phi(y))^2+\alpha(x-sy)^2\right\}\nonumber\\
&=&P(y)\exp\left\{\frac{\alpha}{1-2\alpha}[\phi(y)-sy]^2\right\}\cdot
Q(x|y)\nonumber\\
&\propto&\tilde{P}(y)\cdot
Q(x|y)
\end{eqnarray}
where $Q(x|y)$ is a Gaussian distribution with mean $[\phi(y)-2\alpha
sy]/(1-2\alpha)$ and variance $1/(1-2\alpha)$ and
\begin{equation}
\tilde{P}(y)\propto
P(y)\exp\left\{\frac{\alpha}{1-2\alpha}[\phi(y)-sy]^2\right\}.
\end{equation}
On the other hand, by the first condition of Theorem 1, 
$s$ should be the coefficient pertaining to the optimum
linear estimator of $X$ based on $Y$ under $Q$, which is
\begin{equation}
s=\frac{\bE_Q(XY)}{\bE_Q(Y^2)}=\frac{\bE_Q\{Y\cdot\bE_Q(X|Y)\}}{\bE_Q(Y^2)}.
\end{equation}
But since $Q(x|y)$ is Gaussian with mean $[\phi(y)-2\alpha
sy]/(1-2\alpha)$ as said, then this is exactly the inner expectation at the numerator,
and so, we obtain
\begin{equation}
s=\frac{1}{1-2\alpha}\left[\frac{\bE_Q\{Y\cdot\phi(Y)\}}{\bE_Q(Y^2)}-2\alpha
s\right],
\end{equation}
or equivalently,
\begin{equation}
s=\frac{\bE_Q\{Y\phi(Y)\}}{\bE_Q(Y^2)}.
\end{equation}
But since these expectations involve only the random variable $Y$ whose
marginal under $Q$ is $\tilde{P}$, then the
expectations are actually taken under $\tilde{P}$, i.e.,
\begin{equation}
\label{sequation}
s=\frac{\bE_{\tilde{P}}\{Y\phi(Y)\}}{\bE_{\tilde{P}}(Y^2)}.
\end{equation}
Note that this is different from the solution to the ordinary MMSE
problem, where the solution is given by the same expression, but
with $\tilde{P}$ being replaced by $P$. It should be kept in mind that
$\tilde{P}$ depends, in general, on $s$, then so does the right hand side of
the last equation. We have therefore obtained an equation whose solution
$s=s_Q$ is the optimum coefficient in the sense of minimum
$\bE_P\exp\{\alpha(X-sY)^2\}$. 
Let us now examine a few simple special cases. 

Consider first the case
$\phi(y)=s_0y$, for some real constant $s_0$. In this case, the right
hand--side of eq.\ (\ref{sequation}) is trivially equal to $s_0$, which means
that $s_Q=s_0$. This means that whenever $(X,Y)$ is a Gaussian vector, the
linear MMSE estimator minimizes also
all exponential moments of the squared error (among all linear estimators).

Consider next the case where
\begin{equation}
P(y)=\frac{1}{2}\delta(y-1)+\frac{1}{2}\delta(y+1),
\end{equation}
and denote $\phi_+\eqd\phi(+1)$ and $\phi_-\eqd\phi(-1)$.
Then
\begin{eqnarray}
\tilde{P}(y)&=&\frac{\exp\left\{\frac{\alpha}{1-2\alpha}(\phi_+-s)^2\right\}}
{\exp\left\{\frac{\alpha}{1-2\alpha}(\phi_+-s)^2\right\}+
\exp\left\{\frac{\alpha}{1-2\alpha}(\phi_-+s)^2\right\}}\cdot\delta(y-1)+\nonumber\\
& &\frac{\exp\left\{\frac{\alpha}{1-2\alpha}(\phi_-+s)^2\right\}}
{\exp\left\{\frac{\alpha}{1-2\alpha}(\phi_+-s)^2\right\}+
\exp\left\{\frac{\alpha}{1-2\alpha}(\phi_-+s)^2\right\}}\cdot\delta(y+1).
\end{eqnarray}
Since $\bE_{\tilde{P}}(Y^2)=1$, the equation in $s$ reads
\begin{equation}
s=\frac{\phi_+\exp\left\{\frac{\alpha}{1-2\alpha}(\phi_+-s)^2\right\}-\phi_-
\exp\left\{\frac{\alpha}{1-2\alpha}(\phi_-+s)^2\right\}}
{\exp\left\{\frac{\alpha}{1-2\alpha}(\phi_+-s)^2\right\}+
\exp\left\{\frac{\alpha}{1-2\alpha}(\phi_-+s)^2\right\}}
\end{equation}
For $\phi_+=-\phi_-$, we get $s=\phi_+$, which is expected since this is
actually the linear case discussed above, with $s_0=\phi_+$.
For $\phi_+=\phi_-\eqd\phi$, the equation reads 
\begin{equation}
s=-\phi\tanh\left(\frac{2\alpha\phi s}{1-2\alpha}\right)
\end{equation}
and the only solution is $s=0$, which makes sense since $X$ and $Y$ are
independent in this case. For $\alpha\to 0$, the ordinary MMSE linear
estimator is recovered, whose coefficient is $s=(\phi_+-\phi_-)/2$, as
expected.

Returning to the general setting of this example, let us examine what would
happen if we expand the scope and allow a general, non--linear estimator. In this case, we seek a
general function $s(y)$ such that
\begin{equation}
Q(x,y)\propto P(y)\exp\left\{-\frac{1}{2}(x-\phi(y))^2\right\}\cdot
\exp\{\alpha(x-s(y))^2\}.
\end{equation}
If $\alpha\in(0,1/2)$ and we guess $s(y)=\phi(y)$, we obtain
\begin{equation}
Q(x,y)\propto
P(y)\exp\left\{-\left(\frac{1}{2}-\alpha\right)(x-\phi(y))^2\right\}
\end{equation}
for which the conditional mean, $s_Q(y)=\bE_Q(X|Y=y)$, is indeed $\phi(y)$,
and so, the conditions of Theorem 1 are satisfied. It follows then that 
for our example, $s(y)=\bE(X|Y=y)=\phi(y)$, minimizes not only the MSE,
but also all exponential moments of the squared error,
$\bE\exp\{\alpha(X-s(Y))^2\}$ for $0< \alpha < 1/2$.

The same idea applies to somewhat more general situations. Let $\rho(t)$ be an
even function, which is monotonically non--decreasing for $t\ge 0$, and
steeply enough so that $\int_{-\infty}^{+\infty}\mbox{d}t e^{-\beta\rho(t)} <
\infty$ for all $\beta > \beta_0$, where $\beta_0 > 0$ is a certain constant.
Suppose
that $P(x,y)\propto P(y)\exp[-\beta\rho(x-\phi(y))]$ for some $\beta >
\beta_0$, and we are interested
in minimizing the exponential moment $\bE\exp\{\alpha\rho(X-s(Y))\}$. Then,
for every $\alpha\in(0,\beta-\beta_0)$, the choice $s(y)=\phi(y)$ leaves $Q(x|y)$
symmetric about $x=\phi(y)$. If $\tau=0$ minimizes
$\int_{-\infty}^{+\infty}\mbox{d}t \rho(t-\tau)e^{-\beta\rho(t)}$ for every
$\beta > \beta_0$ (which is true in many cases), then the estimator $s(y)=\phi(y)$ minimizes 
all exponential moments of $\rho(X-s(Y))$. This can be even further
generalized to cases where $P(x|y)\propto \exp[-\beta\rho_1(x-\phi(y))]$ for a
given symmetric function $\rho_1$ that may 
be different from the function $\rho$ for
which we wish to minimize the exponential moment. 

The above considerations extend also to signal estimation (prediction,
filtering, etc.): Consider two jointly wide--sense stationary Gaussian
processes, $\{(X_n,Y_n)\}$. Given $(...,Y_{-1},Y_0,Y_1,\ldots)$, each $X_t$ is
Gaussian, with conditional mean given by
$\bE\{X_t|...,Y_{-1},Y_0,Y_1,\ldots\}=\sum_{i=-\infty}^\infty h_iY_{t-i}$,
$\{h_i\}$ being the impulse response of the non--causal Wiener filter.
From the same reasons as before, 
the exponential moments of the square error are also
minimized by the non--causal Wiener filter. 
It is not clear, however, whether the causal Wiener
filter minimizes the exponentiated square error among all causal filters,
unless the non--causal Wiener filter happens to coincide with the causal
one. Optimum linear prediction of Gaussian processes in the ordinary mean
square error sense are also optimum in the mean exponentiated 
squared error sense.

\subsection{Example 3: Non--Bayesian Estimation}

Let $X_1,X_2,\ldots,X_n$ be i.i.d.\ Gaussian random variables with mean
$\theta$ and variance $\sigma^2$. It is very well known that among all
unbiased estimators of $\theta$, the one the minimizes the mean square error
(or equivalently, the estimation error variance) is the sample mean
$s(x_1,\ldots,x_n)=\frac{1}{n}\sum_{i=1}^n x_i$. Does the sample mean
estimator also minimize
$\bE\exp\{\alpha[s(X_1,\ldots,X_n)-\theta]^2\}$ among all unbiased estimators
and for all values of $\alpha$ in the allowed range?

Once again, the class $\calS$ of all unbiased estimators is clearly a convex set
and $(s-\theta)^2$ is convex in $s$.
Let us `guess' that the sample mean indeed minimizes also
$\bE\exp\{\alpha[s(X_1,\ldots,X_n)-\theta]^2\}$ and then check whether it satisfies
the conditions of Theorem 1. The corresponding probability measure $Q$, which
will be denoted here by $Q_\theta$, is given by
\begin{eqnarray}
Q_\theta(x_1,\ldots,x_n)&\propto&
\exp\left\{-\frac{1}{2\sigma^2}\sum_{i=1}^n(x_i-\theta)^2+
\alpha\left(\frac{1}{n}\sum_{i=1}^nx_i-\theta\right)^2\right\}\nonumber\\
&=&
\exp\left\{-\frac{1}{2\sigma^2}\sum_{i=1}^n(x_i-\theta)^2+
\alpha\left[\frac{1}{n}\sum_{i=1}^n(x_i-\theta)\right]^2\right\}\nonumber\\
&=&
\exp\left\{-\frac{1}{2\sigma^2}(\bx-\theta\bu)^TW(\bx-\theta\bu)\right\},
\end{eqnarray}
where $\bx\eqd(x_1,\ldots,x_n)^T$, $\bu=(1,1,\ldots,1)^T\in\reals^n$ and
\begin{equation}
W=I-\frac{2\alpha\sigma^2}{n^2}\bu\bu^T, 
\end{equation}
$I$ being the $n\times n$ identity
matrix. The maximum likelihood estimator of $\theta$
under $Q_\theta$ is given by
\begin{equation}
s(\bx)=\frac{\bu^TW\bx}{\bu^TW\bu}=\frac{1}{n}\sum_{i=1}^nx_i, 
\end{equation}
namely, the
sample mean. It can easily
be shown to achieve the Cram\'er--Rao
lower bound under $Q_\theta$, which is $\sigma^2/(\bu^TW\bu)$. 
Thus, the sample mean estimator is an optimum unbiased estimator for
$Q_\theta$ and hence it satisfies the conditions of Theorem 1. The answer to
the question of the previous paragraph is then affirmative. The best
achievable performance, in the exponential moment sense, is given by
\begin{equation}
\bE\exp\left\{\alpha\left(\frac{1}{n}\sum_{i=1}^nX_i-\theta\right)^2\right\}
=\frac{1}{\sqrt{\mbox{det}(W)}}.
\end{equation}

\subsection{Example 4: The Gaussian Joint Source--Channel Coding Problem}

Consider the Gaussian memoryless source
\begin{equation}
P_U(\bu)= (2\pi\sigma_u^2)^{-n/2}
\exp\left\{-\frac{1}{2\sigma_u^2}\sum_{i=1}^nu_i^2\right\}
\end{equation}
and the Gaussian memoryless channel $\by=\bx+\bz$, where the noise
is distributed according to
\begin{equation}
P_Z(\bz)=(2\pi\sigma_z^2)^{-n/2}
\exp\left\{-\frac{1}{2\sigma_z^2}\sum_{i=1}^nz_i^2\right\}.
\end{equation}
In the ordinary joint source--channel coding problem, one seeks an encoder and
decoder that would minimize $D=\frac{1}{n}\sum_{i=1}^n\bE\{(U_i-V_i)^2\}$,
where $\bV=(V_1,\ldots,V_n)$ is the reconstruction at the decoder. It is very well
known that the best achievable distortion, in this case, is given by
\begin{equation}
D=\frac{\sigma_u^2}{1+\Gamma/\sigma_z^2},
\end{equation}
where $\Gamma$ is the maximum power allowed at the transmitter, and
it may be achieved by a transmitter that simply amplifies the source
by a gain factor of $\sqrt{\Gamma/\sigma_u^2}$ and a receiver that implements
linear MMSE estimation of $U_i$ given $Y_i$, on a symbol--by--symbol basis.

What happens if we replace the criterion of expected distortion by the
criterion of the exponential moment on the distortion,
$\bE\exp\{\alpha\sum_i(U_i-V_i)^2\}$? It is natural to wonder whether simple
linear transmitters and receivers, of the kind defined in the previous
paragraph, are still optimum.

The random object $X$, in this example, is
the pair of vectors $(\bU,\bZ)$,
where $\bU$ is the source vector and $\bZ$ is the channel noise vector,
which under $P=P_U\times P_Z$, are independent
Gaussian i.i.d.\ random vectors with zero mean and variances
$\sigma_u^2$ and $\sigma_z^2$, respectively, as said. Our strategy $s$
consists of the choice
of an encoding function $\bx=f(\bu)$ and a decoding function $\bv=g(\by)$.
The class $\calS$ is then the set of all pairs of functions $\{f,g\}$, where
$f$ satisfies the power constraint $\bE_P\{\|f(\bU)\|^2\}\le n\Gamma$.
Condition 2 of Theorem 1 tells us that the modified probability distribution
of $\bu$ and $\bz$ should be of the form
\begin{equation}
\label{qjsc}
Q(\bu,\bz)\propto
P_U(\bu)P_Z(\bz)\exp\left\{\alpha\sum_{i=1}^n[u_i-g_i(f(\bu)+\bz)]^2\right\}
\end{equation}
where $g_i$ is restriction of $g$ to the $i$--th component of $\bv$.

Clearly, if we continue to restrict the encoder $f$ to be linear, with a gain of
$\sqrt{\Gamma/\sigma_u^2}$, which simply exploits the allowed power $\Gamma$, and the
only remaining room for 
optimization concerns the decoder $g$, then we are basically
back to the previous example of Bayesian estimation in the Gaussian regime,
and the optimum choice of the decoder 
is a linear one, exactly like in the traditional
mean square error case (from the same consideration as in the Bayesian
estimation example). However, once we extend the scope and allow $f$ to be a non--linear encoder,
then the optimum choice of $f$ and $g$ would no longer remain linear like in
the expected distortion case. It is not difficult to see that the conditions of Theorem 1 are no longer
met for any linear functions $f$ and $g$. The key reason is that while
$Q(\bu,\bz)$ of eq.\ (\ref{qjsc}) continues to be Gaussian (though now $U_i$ and $Z_i$ are
correlated) when $f$ and $g$ are linear, the power constraint,
$\bE_P\{\|\bX\|^2\}\le n\Gamma$, when expressed as an expectation w.r.t.\
$Q$, becomes $\bE_Q\{\|f(\bU)\|^2P(\bU)/Q(\bU)\}\le n\Gamma$, but ``power''
function $\|f(\bu)\|^2P(\bu)/Q(\bu)$, with $P$ and $Q$ being Gaussian densities,
is no longer the usual quadratic function of $f(\bu)$ for
which there is a linear encoder and decoder that is optimum.

Another way to see that linear encoders 
and decoders are suboptimal, is to consider the
following argument: For a given $n$, the expected exponentiated squared error
is minimized by a joint source--channel coding system, defined over a
super-alphabet of $n$--tuples, with respect to a distortion measure, defined
in terms of a single super--letter, as
\begin{equation}
d(\bu,\bv)=\exp\left\{\alpha\sum_{i=1}^n(u_i-v_i)^2\right\}. 
\end{equation}
For such a joint
source--channel coding system to be optimal, the induced channel $P(\bv|\bu)$
must \cite[p.\ 31, eq.\ (2.5.13)]{Berger71} be proportional to 
\begin{equation}
P(\bv)\exp\{-\beta d(\bu,\bv)\}=
P(\bv)\exp\left[-\beta\exp\left\{\alpha\sum_i(u_i-v_i)^2)\right\}\right]
\end{equation}
for some $\beta > 0$,
which is the well--known structure of the optimum test channel that attains the
rate--distortion function for the Gaussian source and the above defined
distortion measure.
Had the aforementioned linear system 
been optimum, the optimum output distribution $P(\bv)$ would be
Gaussian, and then $P(\bv|\bu)$ would remain proportional to a double
exponential function of
$\sum_i(u_i-v_i)^2$. However, the linear system induces instead a Gaussian channel
from $\bu$ to $\bv$, which is very different, and therefore cannot be optimum.

Of course,
the minimum of $\bE\exp\{\alpha\sum_i(U_i-V_i)^2\}$ can be approached by 
separate source- and channel coding, defined on blocks of super--letters
formed by $n$--tuples. The source encoder is an optimum rate--distortion code
for the above defined `single--letter' distortion measure, operating at a
rate close to the channel capacity, and the channel code is 
constructed accordingly to support the same rate.

\section{Universal Asymptotically Optimum Strategies}

The optimum strategy for minimizing
$\bE_P\exp\{\alpha\ell(X,s)\}$ depends, in
general, on both $P$ and $\alpha$.
It turns out, however, that this dependence on $P$ and $\alpha$ can sometimes
be relaxed if one gives up the ambition of deriving a strictly optimum
strategy, and resorts to asymptotically optimum strategies. 

Consider the case where, instead of one random variable $X$, we have a random
vector $\bX=(X_1,\ldots,X_n)$, governed by an product form probability function
\begin{equation}
P(\bx)=\prod_{i=1}^nP(x_i), 
\end{equation}
where each component $x_i$ 
of the vector $\bx=(x_1,\ldots,x_n)$ takes on values in a finite set
$\calX$. If the $\ell(\bx,s)$ grows linearly\footnote{
This happens, for example, when $\ell$ is additive, i.e.,
$\ell(\bx,s)=\sum_{i=1}^nl(x_i,s)$.}
with $n$ for a given empirical
distribution of $\bx$ and a given $s\in\calS$, then it is expected that
the exponential moment $\bE\exp\{\alpha\ell(\bx,s)\}$ would behave, at least
asymptotically, as an exponential function of $n$. In particular, for a given
$s$, the limit 
$$\lim_{n\to\infty}\frac{1}{n}\ln\bE_P\exp\{\alpha\ell(\bX,s)\}$$
exists. Let us denote this limit by $E(s,\alpha,P)$. An {\it asymptotically
optimum} strategy is then a strategy $s^*$ for which 
\begin{equation}
\label{asymopt}
E(s^*,\alpha,P)\le E(s,\alpha,P)
\end{equation}
for every $s\in\calS$. An asymptotically optimum strategy $s^*$ is called
{\it universal asymptotically optimum} w.r.t.\ a class
$\calP$ of probability distributions, if $s^*$ is independent of $\alpha$
and $P$, yet it satisfies eq.\ (\ref{asymopt}) for all $\alpha$
in the allowed range,
every $s\in\calS$, and every $P\in\calP$.
In this section, we take $\calP$ to be the class of all memoryless sources
with a given finite alphabet $\calX$. 
We denote by $T_Q$ the type class pertaining to 
an empirical distribution $Q$, namely, the set of vectors
$\bx\in\calX^n$ whose empirical distribution is $Q$.

Suppose there exists a strategy $s^*$ and a function
$\lambda:\calP\to\reals$ such that following two conditions hold:
\begin{itemize}
\item[(a)] For every type class $T_Q$ and every $\bx\in T_Q$, $\ell(\bx,s^*)\le
n[\lambda(Q)+o(n)]$,
where $o(n)$ designates a (positive) sequence that tends to zero as
$n\to\infty$.
\item[(b)] For every type class $T_Q$ and every $s\in\calS$,
\begin{equation}
\bigg|T_Q\cap\left\{\bx:~\ell(\bx,s)\ge n[\lambda(Q)-o(n)]\right\}\bigg|\ge e^{-no(n)}|T_Q|.
\end{equation}
\end{itemize}
It is then a straightforward exercise to show, using the method of types, that
$s^*$ is a universal asymptotically optimum strategy w.r.t.\ $\calP$, with
\begin{equation}
E(s^*,\alpha,P)=\max_Q[\alpha\lambda(Q)-D(Q\|P)], 
\end{equation}
where condition (a)
supports the direct part and condition (b) supports the converse part. 
The interesting point here then is not quite in the last statement, but in the
fact that there are quite a few application examples where these two
conditions hold at the same time. 

Before we provide such examples, however, a few words are in order concerning
conditions (a) and (b).
Condition (a) means that there is a choice of
$s^*$, that does
{\it not} depend on $\bx$ or on its type class,\footnote{
As before, $s^*$ is chosen without observing the
data first.} yet the performance of $s^*$, for every $\bx\in T_Q$, ``adapts'' to
the empirical distribution $Q$ of $\bx$ in a way, that according to 
condition (b), is ``essentially optimum'' (i.e., cannot be improved
significantly), at least for a
considerable (non--exponential) fraction of the members of $T_Q$.
It is instructive to relate conditions (a) and (b) above to conditions 1 and 2 of
Theorem 1. First, observe that in order to guarantee asymptotic optimality of
$s^*$, condition 2 of Theorem 1 can be somewhat relaxed: For Jensen's
inequality in (\ref{chain}) to remain exponentially tight, it is no longer necessary
to make the random variable $\alpha\ell(\bX,s)+\ln[P(\bX)/Q(\bX)]$ completely
degenerate (i.e., a constant for every realization $\bx$, 
as in condition 2 of Theorem 1), but it is enough to keep it
essentially fixed across a considerably large subset of the
dominant type class,
$T_{Q^*}$, i.e., the one whose empirical distribution 
$Q^*$ essentially achieves the maximum of $[\alpha\lambda(Q)-D(Q\|P)]$.
Taking $Q^*(\bx)$ to be the memoryless source induced by the dominant $Q^*$,
this is indeed precisely what happens under conditions (a) and (b), which
imply that
\begin{eqnarray}
\alpha\ell(\bx,s^*)+\ln\frac{P(\bx)}{Q^*(\bx)}&\approx&
n\alpha\lambda(Q)+\sum_{i=1}^n\ln\frac{P(x_i)}{Q^*(x_i)}\nonumber\\
&=&n\alpha\lambda(Q)+n\sum_{x\in\calX}Q^*(x)\ln\frac{P(x)}{Q^*(x)}\nonumber\\
&=&n[\alpha\lambda(Q^*)-D(Q^*\|P)],
\end{eqnarray} 
for (at least) a non--exponential fraction of the members of $T_{Q^*}$,
namely, a subset of $T_{Q^*}$ 
that is large enough to maintain the exponential order of the
(dominant) contribution of $T_{Q^*}$ to $\bE\exp\{\alpha\ell(\bx,s^*)\}$.
Loosely speaking, the combination of conditions (a) and 
(b) also means then that $s^*$ is essentially
optimum for (this subset of) $T_{Q^*}$, which is a reminiscence of condition 1
of Theorem 1. Moreover, since $s^*$ ``adapts'' to every $T_Q$, in the sense
explained above, then this has the flavor of the max--min 
problem discussed in Section 2, where $s$ is allowed to be optimized for each
and every $Q$. Since the minimizing $s$, in the max-min problem, is independent of $P$ and
$\alpha$, this also explains the universality property of such a strategy.

Let us now discuss a few examples. The first example is that of fixed--rate
rate--distortion coding. A vector $\bX$ that emerges from a memoryless source
$P$ is to be encoded by a coding scheme $s$ 
with respect to a given additive distortion measure, based on a single--letter
distortion measure $d:\calX\times\hat{\calX}\to\reals$, $\hat{\calX}$
being the reconstruction alphabet. Let $D_Q(R)$ denote the
distortion--rate function of a memoryless source $Q$ (with a finite alphabet
$\calX$) relative to the single--letter distortion measure $d$ and
let $\ell(\bx,s)$ designate the distortion between the source vector $\bx$ and
its reproduction, using a rate--distortion code $s$. It is not
difficult to see that this example meets conditions (a) and (b) with
$\lambda(Q)=D_Q(R)$: Condition (a) is based on the type covering lemma
\cite[Section 2.4]{CK81},
according to which each type class $T_Q$ can be completely covered by essentially
$e^{nR}$ `spheres' of radius $nD_Q(R)$ (in the sense of $d$), centered at the
reproduction vectors. Thus $s^*$ can be chosen to be a scheme that encodes
$\bx$ in two parts, the first of which is a header that describes the index of
the type class $T_Q$ of $\bx$ (whose description length is 
proportional to $\log n$) and the second part encodes
the index of the codeword within $T_Q$, using $nR$ nats. Condition (b)
is met since there is no way to cover $T_Q$ with exponentially less than
$e^{nR}$ spheres within distortion less than $D_Q(R)$.

By the same token, consider the dual problem of
variable--rate coding within a maximum allowed distortion $D$. In this case, every source vector
$\bx$ is encoded by $\ell(\bx,s)$ nats, and
this time, conditions (a) and (b) apply with
the choice $\lambda(Q)=R_Q(D)$, which is the rate--distortion function of $Q$
(the inverse function of $D_Q(R)$). The considerations are similar to those of
the first example.
It is interesting to particularize this example, of variable--rate 
coding, to the lossless case, $D=0$ (thus revisiting Example 1), where $R_Q(0)=H_Q$,
the empirical entropy associated with $Q$. In this case, a more refined result can
be obtained, which extends a well known result due to Rissanen
\cite{Rissanen84} in
universal data compression: According to \cite{Rissanen84}, given a length function of
a lossless data compression $\ell(\bx,s)$ ($s$ being the data compression
scheme), and given a parametric class of sources $\{P_\theta\}$, indexed by 
a parameter $\theta\in\Theta\subset\reals^k$, a lower bound on $\bE_\theta\ell(\bX,s)$, 
that applies to most\footnote{``Most values of $\theta$'' means all values of
$\theta$ with the possible exception of a subset of $\Theta$ whose Lebesgue
measure tends to zero as $n$ tends to infinity.} values
of $\theta$, is given by 
\begin{equation}
\bE_\theta \ell(\bX,s)\ge nH_\theta+(1-\epsilon)\frac{k}{2}\log n, 
\end{equation}
where $\epsilon > 0$ is arbitrarily small (for large $n$), $H_\theta$ is the entropy
associated with $P_\theta$, and $\bE_\theta\{\cdot\}$ is the expectation under
$P_\theta$. On the other hand, the same expression is achievable, by a number
of universal coding schemes, provided that
the factor $(1-\epsilon)$ in the above expression is replaced by
$(1+\epsilon)$.
Consider now the case where $\{P_\theta,~\theta\in\Theta\}$ is the
class of all memoryless sources over $\calX$, where the parameter vector $\theta$
designates $k=|\calX|-1$ letter probabilities. As for a lower bound, we have
\begin{eqnarray}
\ln\bE_P\exp\{\alpha\ell(\bX,s)\}&\ge&\max_Q\left[\alpha\bE_Q\ell(\bX,s)-nD(Q\|P)\right]\nonumber\\
&\ge&\max_Q\left\{\alpha\left[nH_Q+(1-\epsilon)\frac{k}{2}\ln
n\right]-nD(Q\|P)\right\}\nonumber\\
&=&n\max_Q\left[\alpha H_Q-D(Q\|P)\right]+\alpha(1-\epsilon)\frac{k}{2}\ln n\nonumber\\
&=&n\alpha H_{1/(1+\alpha)}(P)+\alpha(1-\epsilon)\frac{k}{2}\ln n,
\end{eqnarray}
where the second line follows from Rissanen's lower bound (for most sources),
and where $H_u(P)$ is R\'enyi's entropy of order $u$, namely, 
\begin{equation}
H_u(P)=\frac{1}{1-u}\ln\left[\sum_{x\in\calX}P(x)^u\right].
\end{equation}
Consider now a two--part code $s^*$, which first encodes the index of the type class
$Q$ and then the index of $\bx$ within the type class. The corresponding
length function is given by
\begin{equation}
\ell(\bx,s^*)=\ln|T_Q|+k\ln n\approx n\hat{H}(\bx)+\frac{k}{2}\ln n,
\end{equation}
where $\hat{H}(\bx)$ is the empirical entropy pertaining to $\bx$, and where
the approximate inequality is easily obtained by the Sterling approximation.
Then,
\begin{eqnarray}
\ln\bE_P\exp\{\alpha\ell(\bX,s)\}&=&\ln\bE\exp\{\alpha
n\hat{H}(\bX)\}+\alpha\frac{k}{2}\ln n\nonumber\\
&=&\ln\bE_P\exp\{\alpha \min_Q[-\ln Q(\bX)]\}+\alpha\frac{k}{2}\ln
n\nonumber\\
&\le&\min_Q\ln\bE_P\exp\{-\alpha\ln Q(\bX)\}+\alpha\frac{k}{2}\ln n\nonumber\\
&=&n\alpha H_{1/(1+\alpha)}(P)+\alpha\frac{k}{2}\ln n,
\end{eqnarray}
and then it essentially achieves the lower bound. 
Rissanen's result is now a special case
of this, corresponding to $\alpha\to 0$.

Our last example corresponds to a secrecy system. A sequence $\bx$ is to be
communicated to a legitimate decoder which shares with the transmitter a
random key $\bz$ of $nR$ purely random bits. The encoder transmits an
encrypted message $\by=\phi(\bx,\bz)$, which is an invertible function of
$\bx$ given $\bz$, and hence decipherable by the legitimate decoder. An
eavesdropper, which has no access to the key $\bz$, submits a sequence of guesses
concerning $\bx$ until it receives an indication that the last guess was
correct (e.g., a correct guess of a password admits the eavesdropper into a secret
system). For the best possible encryption function $\phi$, what would be the
optimum guessing strategy $s^*$ that the eavesdropper may apply 
in order to minimize the $\alpha$--th moment of the number of
guesses $G(\bX,s)$, i.e., $\bE\{G^\alpha(\bX,s)\}$? In this case,
$\ell(\bx,s)=\ln G(\bx,s)$. As is shown in \cite{MA99}, there exists a
guessing strategy $s^*$, which for every $\bx\in T_Q$, gives
$\ell(\bx,s^*)\approx n\min\{H_Q,R\}$, a quantity that essentially cannot be 
improved upon by any other guessing strategy, for most members of $T_Q$. In other
words, conditions (a) and (b) apply with $\lambda(Q)=\min\{H_Q,R\}$. 

\section{Phase Transitions}

Another interesting aspect of the asymptotic behavior of the exponential
moment is the possible appearance of phase transitions, i.e., 
irregularities in the exponent function
$E(s,\alpha,P)$ even in some very simple and `innocent' models. 
By irregularities, we mean a non--smooth behavior, namely,
discontinuities in the derivatives of $E(s,\alpha,P)$ with respect to $\alpha$
and/or the parameters of the source $P$. 

One example that exhibits phase 
transitions is that of the secrecy system,
mentioned in the last paragraph of the previous section. As is shown in
\cite{MA99}, the optimum exponent $E(s^*,\alpha,p)$ for this case
consists of two phase transitions as a function of $R$ (namely, three different
phases). In particular,
\begin{equation}
E(s^*,\alpha,P)=\left\{ \begin{array}{ll}
\alpha R & R < H(P) \\
(\alpha-\theta_R)R+\theta_R H_{1/(1+\theta_R)}(P) & H(P)\le R\le H(P_\alpha) \\
\alpha H_{1/(1+\alpha)}(P) & R > H(P_\alpha)
\end{array} \right.
\end{equation}
where $P_\alpha$ is the distribution defined by
\begin{equation}
P_\alpha(x)\eqd\frac{P^{1/(1+\alpha)}(x)}{\sum_{x'\in\calX} P^{1/(1+\alpha)}(x')},
\end{equation}
$H(Q)$ is the Shannon entropy associated with a distribution $Q$, 
$H_u(Q)$ is the R\'enyi entropy of order $u$ as defined before, and
$\theta_R$ is the unique solution of
the equation $R=H(P_\theta)$ for $R$ in the range
$H(P) \le R \le H(P_\alpha)$.
But this example may not really be extremely 
surprising due to the non--smoothness of the function
$\lambda(Q)=\min\{H_Q,R\}$.

It may be somewhat less expected, however, to witness phase
transitions also in some very simple and `innocent' looking models.
One way to understand the
phase transitions in these cases,
comes from the statistical--mechanical perspective. It turns
out that in some cases, the expression of the exponential moment is analogous
to that of a partition function of a certain many--particle physical
system with interactions, which may exhibit phase transitions and these phase
transitions correspond the above--mentioned irregularities. 

We now demonstrate a very simple model, which has phase transitions.
Consider the case where $\bX$ is a binary vector whose components take
on values in $\calX=\{-1,+1\}$, and which is governed by a binary memoryless
source $P_\mu$ with probabilities
$\mbox{Pr}\{X_i=+1\}=1-\mbox{Pr}\{X_i=-1\}=(1+\mu)/2$ ($\mu$ designating the
expected `magnetization' of each binary spin $X_i$, to make the physical
analogy apparent). The probability of $\bx$ under $P_\mu$ is thus easily
shown to be given by
\begin{equation}
P_\mu(\bx)=\left(\frac{1+\mu}{2}\right)^{(n+\sum_ix_i)/2}\cdot
\left(\frac{1-\mu}{2}\right)^{(n-\sum_ix_i)/2}
=\left(\frac{1-\mu^2}{4}\right)^{n/2}\cdot\left(\frac{1+\mu}{1-\mu}\right)^{\sum_ix_i/2}.
\end{equation}
Consider the estimation of the parameter $\mu$ by the ML estimator
\begin{equation}
\hat{\mu}=\frac{1}{n}\sum_{i=1}^nx_i.
\end{equation}
How does the exponential moment of $\bE_\mu\exp\{\alpha n(\hat{\mu}-\mu)^2\}$
behave like? A straightforward derivation yields
\begin{eqnarray}
\bE_\mu\exp\{\alpha n(\hat{\mu}-\mu)^2\}&=&
\left(\frac{1-\mu^2}{4}\right)^{n/2}e^{n\alpha\mu^2}
\sum_{\bx}\left(\frac{1+\mu}{1-\mu}\right)^{\sum_ix_i/2}\exp\left\{\frac{\alpha}{n}\left(\sum_ix_i\right)^2
-2\alpha\mu\sum_ix_i\right\}\nonumber\\
&=&\left(\frac{1-\mu^2}{4}\right)^{n/2} e^{n\alpha\mu^2}
\sum_{\bx}\exp\left\{\left(\frac{1}{2}\ln\frac{1+\mu}{1-\mu}-2\alpha\mu\right)\sum_ix_i+
\frac{\alpha}{n}\left(\sum_ix_i\right)^2\right\}.\nonumber
\end{eqnarray}
The last summation over $\{\bx\}$ is exactly the partition function pertaining
to the {\it Curie--Weiss model} of spin arrays in statistical mechanics (see, e.g.,
\cite[Subsection 2.5.2]{MM09}), where
the magnetic field is given by 
\begin{equation}
B=\frac{1}{2}\ln\frac{1+\mu}{1-\mu}-2\alpha\mu
\end{equation}
and the coupling coefficient for every pair of spins is $J=2\alpha$. 
It is well known that this model exhibits phase transitions pertaining to
spontaneous magnetization below a certain critical temperature. In particular,
using the method of types \cite{CK81}, this partition function can be asymptotically evaluated
as being of the exponential order of
$$\exp\left\{n\cdot\max_{|m|\le
1}\left[h_2\left(\frac{1+m}{2}\right)+Bm+\frac{J}{2}\cdot m^2\right]\right\},$$
where $h_2(\cdot)$ is the binary entropy function, which stands for the
exponential order of the number of configurations $\{\bx\}$ with a given
value of $m=\frac{1}{n}\sum_ix_i$. This expression
is clearly dominated by a
value of $m$ (the dominant magnetization $m^*$)
which maximizes the expression in the square brackets, i.e., it solves the equation  
\begin{equation}
m=\tanh(Jm+B),
\end{equation}
or in our variables,
\begin{equation}
m=\tanh\left(2\alpha m+\frac{1}{2}\ln\frac{1+\mu}{1-\mu}-2\alpha\mu\right).
\end{equation}
For $\alpha < 1/2$, there is only one solution and there is no spontaneous
magnetization (paramagnetic phase). For $\alpha > 1/2$, however, there are
three solutions, and only one of them dominates the partition function,
depending on the sign of $B$, or equivalently, on whether $\alpha >
\alpha_0(\mu)\eqd\frac{1}{4\mu}\ln\frac{1+\mu}{1-\mu}$ or $\alpha <
\alpha_0(\mu)$ and according to the sign of $\mu$. Accordingly, there are five
different phases in the plane spanned by $\alpha$ and $\mu$. The paramagnetic
phase $\alpha < 1/2$, the phases $\{\mu > 0,~1/2 < \alpha < \alpha_0(\mu)\}$
and $\{\mu < 0,~ \alpha > \alpha_0(\mu)\}$, where the dominant magnetization
$m$ is positive, and the two complementary phases,
$\{\mu < 0,~1/2 < \alpha < \alpha_0(\mu)\}$ and
and $\{\mu > 0,~ \alpha > \alpha_0(\mu)\}$, where the dominant magnetization
is negative. Thus, there is a multi-critical point 
where the boundaries of all five phases meet, which the point 
$(\mu,\alpha)=(0,1/2)$. The phase diagram is depicted in Fig.\ 1.

\begin{figure}[ht]
\hspace*{3cm}\input{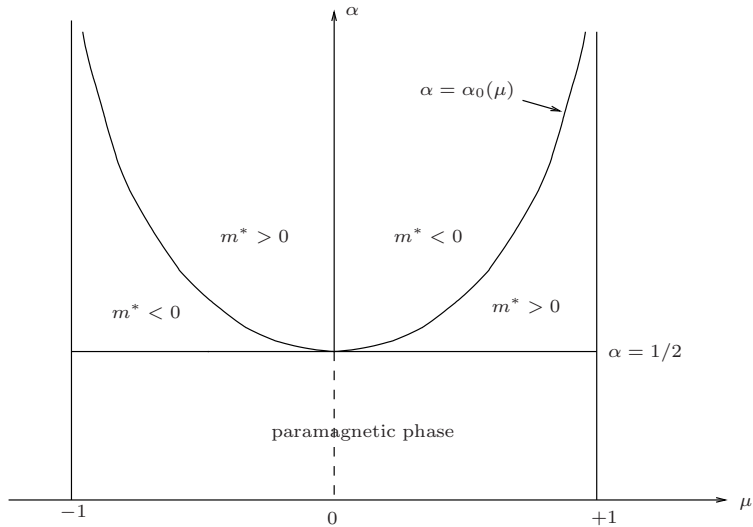}
\caption{\small Phase diagram in the plane of $(\mu,\alpha)$.}
\label{mm1}
\end{figure}

Yet another example of phase transitions is that of fixed--rate lossy data
compression, discussed in the previous section.
To demonstrate this explicitly, consider the binary symmetric
source (BSS) and the Hamming distortion measure $d$, and consider a random
selection of a rate--$R$ code
by $ne^{nR}$ independent fair coin tosses, one for each of the $n$ components
of every one of the $e^{nR}$ codewords. It was shown in \cite{Merhav09} that
the asymptotic exponent of the negative exponential moment,
$\bE\exp\{-\alpha\sum_id(U_i,V_i)\}$ (where
the expectation is w.r.t.\ both the source and the random code selection),
is given by the following expression, which obviously exhibits a (second
order) phase transition:
\begin{equation}
\lim_{n\to\infty}\frac{1}{n}\ln
\bE\exp\left\{-\alpha\sum_id(U_i,V_i)\right\} 
=\left\{\begin{array}{ll}
-\alpha\delta (R) & \alpha\le \alpha(R)\\
-\alpha+\ln(1+e^\alpha)+R-\ln 2 & \alpha > \alpha(R)
\end{array}
\right.
\end{equation}
where $\delta(R)$ is the distortion--rate function of the BSS w.r.t.\
the Hamming distortion measure and
\begin{equation}
\alpha(R)=\ln\frac{1-\delta(R)}{\delta(R)}.
\end{equation}
The analysis in \cite{Merhav09} is based on the random energy model (REM),
\cite{Derrida80},\cite{Derrida80b},\cite{Derrida81}, 
a well-known statistical--mechanical model of spin glasses with
strong disorder, which is known to exhibit phase transitions. Moreover, it is
shown in \cite{Merhav09} that ensembles of codes that have an hierarchical
structure may have more than one phase transition.

\section{Lower Bounds on Exponential Moments}

As explained in the Introduction,
even in the ordinary setting, of the quest for minimizing
$\bE\{\ell(X,s)\}$, optimum strategies may not always be known, and then
useful lower bounds are very important. 
This is definitely the case when exponential moments are 
considered, because the exponential moment criterion is even harder to handle. To obtain non--trivial
bounds on exponential moments, we propose to harness lower bounds the expectation of $\ell(X,s)$,
possibly using a change of measure, in the spirit of the proof of Theorem 1
and the previous example of a lower bound on universal lossless data
compression. We next demonstrate this idea in the context of a lower bound on the expected
exponentiated squared error of an unbiased estimator, on the basis of the Cram\'er--Rao bound (CRB).
The basic idea, however, is applicable more generally, e.g., by relying on 
other well--known Bayesian/non--Bayesian bounds on the mean-square error (e.g., the Weiss--Weinstein
bound for Bayesian estimation \cite{WW85}), as well as in
bounds on signal estimation (filtering, prediction, etc.),
and in other problem areas as well. Further investigation in the line may be of
considerable interest.

Consider a parametric family of probability distributions
$\{P_\theta,~\theta\in\Theta\}$, $\Theta\subseteq\reals$ 
being the parameter set, and suppose that we are interested in
a lower bound on $\bE_\theta\exp\{\alpha(\hat{\theta}-\theta)^2\}$, for
any unbiased estimator of $\theta$, where as before, $\bE_\theta$ denotes
expectation w.r.t.\ $P_\theta$. Consider the following chain of inequalities,
which holds for any $\theta'\in\Theta$:
\begin{eqnarray}
\bE_\theta\exp\{\alpha(\hat{\theta}-\theta)^2\}&=&
\bE_{\theta'}\exp\left\{\alpha(\hat{\theta}-\theta)^2+\ln\frac{P_\theta(X)}{P_{\theta'}(X)}\right\}\nonumber\\
&\ge&\exp\left\{\alpha\bE_{\theta'}(\hat{\theta}-\theta)^2-D(P_{\theta'}\|P_\theta)\right\}\nonumber\\
&=&\exp\left\{\alpha\bE_{\theta'}(\hat{\theta}-\theta')^2+
\alpha(\theta-\theta')^2-D(P_{\theta'}\|P_\theta)\right\}\nonumber\\
&\ge&\exp\left\{\alpha\mbox{CRB}(\theta')+
\alpha(\theta-\theta')^2-D(P_{\theta'}\|P_\theta)\right\},
\end{eqnarray}
where $\mbox{CRB}(\theta)$ is the Cram\'er--Rao bound for unbiased estimators, computed at $\theta$ (i.e.,
$\mbox{CRB}(\theta)=1/I(\theta)$, where $I(\theta)$ is the Fisher information). Since
this lower bound applies for every $\theta'\in\Theta$, one can take its
supremum over $\theta'\in\Theta$ and obtain
\begin{equation}
\ln\bE_\theta\exp\{\alpha(\hat{\theta}-\theta)^2\}\ge
\sup_{\theta'\in\Theta}\left[\alpha\mbox{CRB}(\theta')+\alpha(\theta'-\theta)^2-D(P_{\theta'}\|P_\theta)\right].
\end{equation}
More generally if $\theta=(\theta_1,\ldots,\theta_k)^T$ 
is a parameter vector (thus
$\theta\in\Theta\subseteq\reals^k$) and $\alpha\in\reals^k$ is an arbitrary
deterministic (column) vector, then
\begin{equation}
\ln\bE_\theta\exp\{\alpha^T(\hat{\theta}-\theta)(\hat{\theta}-\theta)^T\alpha\}\ge
\sup_{\theta'\in\Theta}\left[\alpha^TI^{-1}(\theta')\alpha+[\alpha^T(\theta'-\theta)]^2-
D(P_{\theta'}\|P_\theta)\right],
\end{equation}
where here $I(\theta)$ is the Fisher information matrix and $I^{-1}(\theta)$
is its inverse.

It would be interesting to further investigate bounds of this type, in
parameter estimation in particular, and in other problem areas in general,
and to examine when these bounds may be tight and useful.

\section*{Acknowledgment}

Interesting discussions with Rami Atar are acknowledged with thanks.




\begin{thebibliography}{AA}

\bibitem{Arikan96}
E.~Arikan,
``An inequality on guessing and its application to sequential decoding,''
{\sl IEEE Trans.\ Inform.\ Theory,} vol.\ IT--42, no.\ 1, pp.\ 99--105, January 1996.

\bibitem{AM98a}
E.~Arikan and N.~Merhav, ``Guessing subject to distortion,''
{\it IEEE Trans.\ Inform.\ Theory}, vol.\ 44,
no.\ 3, pp.\ 1041--1056, May 1998.

\bibitem{AM98b}
E.~Arikan and N.~Merhav, ``Joint source--channel coding and
guessing with application to sequential decoding,''
{\it IEEE Trans.\ Inform.\ Theory}, vol.\ 44, no.\ 5, pp.\ 1756--1769,
September 1998.

\bibitem{Berger71}
T.~Berger, {\it Rate Distortion Theory: A Mathematical Basis for Data
Compression}, Prentice Hall, Englewood Cliffs, New Jersey, U.S.A., 1971.

\bibitem{Bertsekas07}
D.~Bertsekas, {\it Dynamic Programming and Optimal Control}, Vol.\ I and II,
3rd ed.\ Nashua, NH: Athena Scientific, 2007.

\bibitem{CT06}
T.~M.~Cover and J.~A.~Thomas, {\it Elements of Information Theory}, Second
Edition, John Wiley \& Sons, Hoboken, New Jersey, U.S.A., 2006. 

\bibitem{CK81}
I.~Csisz\'ar and J.~K\"orner, {\it Information Theory: Coding Theorems for
Discrete Memoryless Systems}, Academic Press, New York, U.S.A., 1981.

\bibitem{Derrida80}
B.~Derrida, ``Random--energy model: limit of a family of disordered models,''
{\it Phys.\ Rev.\ Lett.}, vol.\ 45, no.\ 2, pp.\ 79--82, July 1980.

\bibitem{Derrida80b}
B.~Derrida, ``The random energy model,'' {\it Physics Reports} (Review Section
of Physics Letters), vol.\ 67, no.\ 1, pp.\ 29--35, 1980.

\bibitem{Derrida81}
B.~Derrida, ``Random--energy model: an exactly solvable model for disordered
systems,'' 
{\it Phys.\ Rev.\ B}, vol.\ 24, no.\ 5, pp.\ 2613--2626, September 1981.

\bibitem{FM86}
N.~Farvardin and J.~W.~Modestino, ``On overflow and underflow problems
in buffer instrumented variable-length coding of fixed-rate memoryless
sources,''
{\em IEEE Transactions on Information Theory\/},
vol.~IT--32, no.~6, pp.~839--845, November 1986.

\bibitem{Humblet81}
P.~A.~Humblet, ``Generalization of Huffman coding to minimize the 
probability of buffer overflow,''
{\em IEEE Transactions on Information Theory\/},
vol.~IT--27, no.~2, pp.~230--232, March 1981.

\bibitem{Jelinek68}
F.~Jelinek, ``Buffer overflow in variable length coding of fixed 
rate sources,''
{\em IEEE Transactions on Information Theory\/},
vol.~IT--14, no.~3, pp.~490--501, May 1968.

\bibitem{Massey94}
J.~L.~Massey, ``Guessing and entropy,'' {\it Proc.\ 1994 IEEE Int.\ Symp.\
Inform.\ Theory}, (ISIT `94), p.\ 204, Trondheim, Norway, 1994.

\bibitem{Merhav91}
N.\ Merhav, ``Universal coding with
minimum probability of code word length
overflow,''
{\em IEEE Trans. Inform. Theory},
vol.~37, no.~3, pp.~556--563, May 1991.

\bibitem{Merhav09}
N.~Merhav, ``The generalized random energy model and its
application to the statistical physics of
ensembles of hierarchical codes,''
{\it IEEE Trans.\ Inform.\ Theory}, vol.\ 55, no.\ 3, pp.\ 1250--1268, March
2009.

\bibitem{MA99}
N.~Merhav and E.~Arikan, ``The Shannon
cipher system with a guessing wiretapper,''
{\it IEEE Trans.\ Inform.\ Theory}, vol.\ 45,
no.\ 6, pp.\ 1860--1866, September 1999.

\bibitem{MF98}
N.~Merhav and M.~Feder, ``Universal prediction,'' 
{\it IEEE Trans.\ Inform.\
Theory}, vol.\ 44, no.\ 6, pp.\ 2124--2147, October 1998.

\bibitem{MN92}
N.~Merhav and D.~L.~Neuhoff, ``Variable-to-fixed length codes provide
better large deviations performance than fixed-to-variable length codes,''
{\em IEEE Trans.\ Inform.\ Theory},
vol.~38, no.~1, pp.~135--140, January 1992.

\bibitem{MRA99}
N.~Merhav, R.~M.~Roth, and E.~Arikan, ``Hierarchical guessing with a
fidelity criterion,''
{\it IEEE Trans.\ Inform.\ Theory}, vol.\ 45,
no.\ 1, pp.\ 330--337, January 1999.

\bibitem{MM09}
M.~M\'ezard and
A.~Montanari, {\it Information, Physics and Computation},
Oxford University Press, 2009.

\bibitem{Rissanen84}
J.~Rissanen, ``Universal coding, information, prediction, and 
estimation,'' {\em IEEE Transactions on Information Theory\/},
vol.~IT--30, no.~4, pp.~629--636, July 1984.

\bibitem{UH99}
O.~Uchida and T.~S.~Han, ``The optimal overflow and underflow 
probabilities with variable--length coding for the general source,''
preprint 1999.

\bibitem{VanTrees68}
H.~van Trees, {\it Detection, Estimation, and Modulation Theory}, Part I,
John Wiley \& Sons, New York, 1968. 

\bibitem{WW85}
A.~J.~Weiss and E.~Weinstein, ``A lower bound on the mean square error
in random parameter estimation,''
{\em IEEE Trans.\ Inform.\ Theory\/},
vol.~IT--31, no.~5, pp.~680--682, September 1985.

\bibitem{Wyner74-1}
A.~D.~Wyner, ``On the probability of buffer overflow under an arbitrary
bounded input-output distribution,''
{\em SIAM Journal on Applied Mathematics\/},
vol.~27, no.~4, pp.~544--570, December 1974.

\bibitem{Wyner80}
A.~D.~Wyner, ``Another look at the coding theorem of information theory-
a tutorial,''
{\em Proc. IEEE\/}, vol.~58, no.~6, pp.~894--913, June 1980.

\bibitem{Wyner81}
A.~D.~Wyner, ``Fundamental limits in information theory,'' 
{\em Proc. IEEE\/}, vol.~69, no.~2, pp.~239--251, February 1981.

\end{thebibliography}
\end{document}